\documentclass[reprint,aps]{revtex4-2}

\usepackage{graphicx}
\usepackage{amsmath}
\usepackage{dcolumn}
\usepackage{bm}
\usepackage{color}
\usepackage{pifont}
\usepackage{epstopdf}
\usepackage{booktabs}

\makeatletter

\newcommand{\Rmnum}[1]{\expandafter\@slowromancap\romannumeral #1@}
\makeatother

\begin{document}
	\title{Impact of individual behavioral changes on epidemic spreading in time-varying networks}
	\author{Bing Wang$^1$}
	\author{Zeyang Xie$^1$}
	\author{Yuexing Han$^{1,2}$}
	\email[]{han$_$yx@i.shu.edu.cn }
	\affiliation{$^1$School of Computer Engineering and Science, Shanghai University, Shanghai, P.R. China}
	\affiliation{$^2$Shanghai Institute for Advanced Communication and Data Science, Shanghai University, Shanghai, P.R. China}
	\date{\today}
	\begin{abstract}
	Changes in individual behavior often entangles with the dynamic interaction of individuals, which complicates the epidemic process and brings great challenges for the understanding and control of the epidemic. In this work, we consider three kinds of typical behavioral changes in epidemic process that is, self-quarantine of infected individuals, self-protection of susceptible individuals, and social distancing between them. We connect the behavioral changes with individual's social attributes by the activity-driven network with attractiveness. A mean-field theory is established to derive an analytical estimate of epidemic threshold for SIS models with individual behavioral changes, which depends on the correlations between activity, attractiveness and the number of generative links in the susceptible and infected states. We find that individual behaviors play different roles in suppressing the epidemic. Although all the behavioral changes could delay the epidemic by increasing the epidemic threshold, self-quarantine and social distancing of infected individuals could effectively decrease the epidemic size. In addition, simultaneous changes in these behaviors and the timing of implement of them also play a key role in suppressing the epidemic. These results provide helpful significance for understanding the interaction of individual behaviors in the epidemic process.
	\end{abstract}
	\maketitle

	\section{Introduction}~\label{sec:Intro}
	Faced with the epidemic, such as SARS, H1N1 influenza and the ongoing Coronavirus disease 2019(COVID-19), non-pharmaceutic measures are often adopted to halt epidemic propagation \cite{Lai2020,Kraemer2020}. On general, there are two types of non-pharmaceutic measures \cite{Gozzi2020}. The first describes top-down measures, such as lock down of a city \cite{Carcione2020,ShiWenzhong2020}, school closure \cite{Lee2020}, travel restrictions \cite{Tian2020,Jia2020}, contact tracing \cite{Ferretti2020,Walker2020}, mandatory quarantine \cite{Li2020,Victor2020}, taken by governments aimed at interrupting the chain of infection. The second is bottom-up measures driven by individual behavioral changes, such as self-quarantine \cite{Ryu2020}, social distancing \cite{Huang2021}, washing hands frequently and wearing face masks \cite{Eikenberry2020}. As the epidemic gets under control, top-down measures are gradually lifted to reduce economic risk and social problems (e.g. unemployment, educational inefficiency) \cite{Hsiang2020,Chowdhury2020}, while bottom-up individual behavioral changes still play a key role in curbing the epidemic \cite{Gozzi2020}. Thus, studying the effects of bottom-up individual behavioral changes on epidemic dynamics is essential to curb epidemic spreading.
	
	Many researchers have focused their attention on this subject and provided a wealthy of theoretical models to capture the effects of individual behavioral changes on epidemic. In a few studies \cite{Kraemer2020,Perra2011,Meloni2011}, individual behavioral changes are implemented after the initial growth of infected individuals. For instance, Eikenberry et al. \cite{Eikenberry2020} investigated the potential for face masks used by public to curtail the COVID-19 and found that the use of face masks could reduce the epidemic size. Lin et al. \cite{Lin2020} considered the decreasing contacts among individuals responding to the number of deaths in COVID-19, and modeled it as the variations in infection rate. It reveals that as the increase of the individual reaction's intensity, the epidemic size gradually decreases. Indeed, prevalence-induced behavioral changes do not affect epidemic threshold, but can drastically reduce the final epidemic size \cite{Gozzi2020}. This kind of approaches are further refined by extending original epidemiological models to include behavior-related compartments, such as vaccine or quarantine \cite{Kabir2019,Feng2011,Fukuda2016}, in which individual's susceptibility is reduced by some parameters. Thus, the extension of epidemiological models also could modify the conditions for epidemic outbreak \cite{Huang2021,Kang2017}. In all, most of previous studies focus on the behavioral changes by assuming that the interaction of individual is static, which may lead to incorrect conclusions \cite{Valdano2015,Holme2014,Liu2014}.
	
	 In reality, interactions between individuals usually dynamically evolve, and is affected by individual behavioral changes. Such changes driven by individuals further alter the epidemic dynamics. Some studies use adaptive networks to investigate the impact of individual behavioral changes on epidemic dynamics \cite{Epstein2008,Shaw2008}. Gross et al \cite{Gross2006} proposed a rewiring mechanism to mimic the fact that susceptible individuals might be aware of the epidemiological state of their neighbors, and  actively avoid being connected with the infected. Guo et al \cite{Guo2013} proposed a similar model in which the contacts between infected and susceptible individuals are deactivated. Other studies considered the variations in individuals' social attributes to reflect their behavioral changes \cite{Mancastroppa2020,Zino2019,Zino2020}. For example, based on activity-driven model(AD), Gozzi et al \cite{Gozzi2020} considered that a fraction of people reduce their social interactions after experiencing a first wave of infections and found that reduction in the number of interactions of a small number of highly active individuals could effectively curb the disease. Considering persistent contacts between individuals, Nadini et al \cite{Nadini2020} proposed temporal networks with static backbone where adaptive behaviors related with epidemiological states alter individuals' activity. It shows that the existence of persistent links will form local communities of infected individuals. Although the impact of individual behavioral changes on the epidemic dynamics has been widely explored on the time-varying networks, most of studies focus only on one kind of individual behavior and ignore the joint effects of them.
			
	 In this paper, we investigate the effects of joint changes of typical behaviors on epidemic unfolding on time-varying networks. Usually, once infected individuals are detected and quarantined, they may actively reduce the chance to attend social activities, thus we name it as self-quarantine. Simultaneously, healthy individuals also reduce the contact with infected individuals to avoid infection, and we name this behavior as self-protection. In addition, social distancing driven by infected individuals to reduce the contact between them and other individuals is also considered. Based on the Susceptible-Infected-Susceptible(SIS) epidemic process, we explore the interaction between social attributes such as individual's activity, attractiveness, the number of contacts and behavioral changes caused by the epidemic dynamic, such as self-quarantine, self-protection, social distancing, on the activity-driven network with attractiveness(ADA) model. Specifically, self-quarantine and social distancing of infected individuals reduce their activity and number of generative links, respectively, while the self-protection behavior of susceptible individuals reduce infected individuals' attractiveness. We adopt the mean-field theory to evaluate the impacts of behavioral changes on the epidemic. Through the Monte Carlo experiments in synthetic networks, we verify the theoretical epidemic threshold. We find that all the behavioral changes that we testified could delay the epidemic by increasing the epidemic threshold, while the self-quarantine and social distancing of infected individuals can efficiently decrease the epidemic size. The joint changes of the three behaviors could further curb the epidemic. In addition, the timing that individuals change their behavior also plays a fundamental role in suppressing the epidemic.
	
	The remainder of this paper is organized as follows. In Section.\ref{sec:Methods}, we introduce the ADA model and the behavioral changes with SIS epidemic process. In Section.\ref{sec:theory}, we derive the epidemic threshold in detail. In Section.\ref{sec:SimuonSyn}, we present experimental results of the proposed model and compare the effect of different behavioral changes on disease dynamics. In Section.\ref{sec:SimuonReal}, we show the result of the proposed model on real network, and conclusions are presented in Section.\ref{sec.conclu}.
	
	\section{Model description}~\label{sec:Methods}
	On the ADA model, a network composed of $N$ nodes is characterized by a joint probability distribution $F(a,b)$, from which the activity $a_i$ and attractiveness $b_i$ of each node $i$ are extracted \cite{Pozzana2017}. For the convenience of analysis, we assume activity and attractiveness follow an independent power law distribution, that is  $F(a)\propto  a^{-\gamma_a}$($a\in[\epsilon,a_{max}]$) and $F(b)\propto b^{-\gamma_b}$($b\in[\epsilon,b_{max}]$) with $2\leq \gamma_a,\gamma_b \leq 3$ and cut-offs $\epsilon$ to avoid divergences \cite{Rizzo2016,Perra2012}. The activity and attractiveness of each node $i$ are expressed as the vector ($a_i$,$b_i$). Thus, node $i$ activates with probability $a_i \Delta T$ and then connects one of $m$ links towards node $j$ with probability proportional to its attractiveness $b_j$.
	
	\begin{figure}
		\includegraphics[scale=0.45]{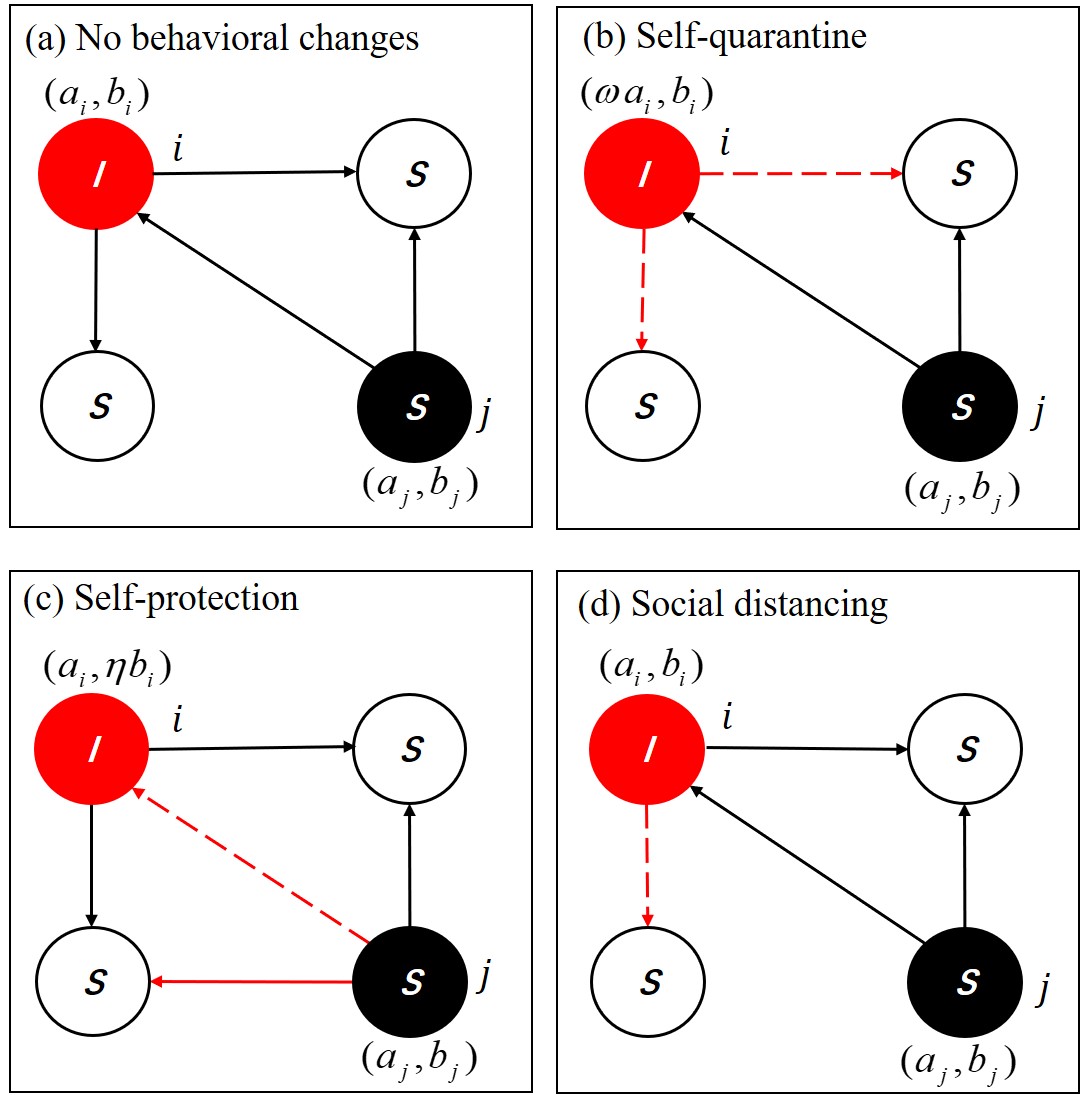}
		\caption{Behavioral changes on the ADA model using SIS epidemic processes. Red and black filled circles represent actively infected and susceptible individuals, respectively, and empty circles represent inactive individuals. (a) No behavioral changes; (b) Self-quarantine of infected individuals by reducing their activity; (c) Self-protection of susceptible individuals by reducing infected individuals' attractiveness; (d) Social distancing of infected individuals by reducing the number of generative links.}
		\label{fig:individualBehavior}
	\end{figure}

 	 Using SIS processes on the ADA model, the activity and attractiveness of each susceptible individual $j$ are described as the vector $(a_{j,S},b_{j,S})=(a_j,b_j)$, and the number of generative links per activation is $m_S=m$. With the introduction of behavioral changes, infected individuals' activity, attractiveness and number of generative links per activation will change. We consider three kinds of behavioral changes including self-quarantine and social distancing of infected individuals, self-protection of susceptible individuals, as shown in Fig.\ref{fig:individualBehavior}. Self-quarantine of infected individuals means that they reduce the chance of participating in social activities, due to the appearance of symptoms. Thus, an infected individual $i$ reduces his activity by factor $\omega$ ($\omega \in (0,1]$), i.e., $\omega a_i$. Self-protection of susceptible individuals refers that they are conscious to avoid the contact with infected individuals, which leads to the reduction in attractiveness of an infected individual $i$ by factor $\eta$ ($\eta \in (0,1]$), i.e., $\eta b_i$. Thus, the activity and attractiveness of each infected individual $i$ are updated as the vector $(a_{i,I},b_{i,I})=(\omega a_i,\eta b_i)$. Social distancing usually refer to the reduction in number of contacts between individuals to prevent epidemic spreading. Here, we assume that social distancing of infected individuals indicates the reduction in number of contacts between them and other individuals, so the number of generative links per activation for infected individuals is reduced to $c_m m$. 
	
	The generative process of instantaneous network $G_t$ is similar to the process presented in \cite{Pozzana2017}. With the behavioral changes being introduced, the process is described as follows: 
	\begin{itemize}
		\item According to the given activity distribution $F(a)$ and attractiveness distribution $F(b)$, each node $i$ is allocated activity $a_i$ and attractiveness $b_i$;
		\item At each discrete time $t$, each node $i$ in state $X \in\left\lbrace S,I\right\rbrace $ becomes active with probability $a_{i,X}\Delta t$ and creates $m_{X}$ links with other nodes in the instantaneous network $G_t$. Specifically, active node $i$ chooses node $j$ as a target of one of its connections according to the probability proportional to the attractiveness $b_{j,X}$ of node $j$ in state $X$;
		\item The epidemic dynamics takes place over the instantaneous network $G_t$;
		\item At the next time step $t+\Delta t$, each link on the instantaneous network $G_t$ is removed, and the process resumes from the second point.
	\end{itemize}
	\section{Epidemic Threshold}~\label{sec:theory}
	 We model the spreading of an infectious disease unfolding at a comparable time-scale with respect to the evolution of connections on the contact network. Without loss of generality, we set $\Delta t=1$. With the SIS epidemic processes, each individual can be in one of two states: susceptible or infected. A susceptible individual will get infected by contacting with an infected individual with rate $\lambda$. The infected individual recovers spontaneously with rate $\mu$. In the absence of behavioral changes, by studying the evolution of the number of infected individuals in class $(a,b)$, the epidemic threshold on the original ADA model can be obtained as follows:
	 \begin{equation}
	 	~\label{eqn:ADAThreshold}
	 	\frac{\lambda}{\mu}=\frac{1}{m}\frac{2 \left\langle b\right\rangle }{\left\langle ab\right\rangle +\sqrt{\left\langle a^2\right\rangle \left\langle b^2\right\rangle }}\text{,}
	 \end{equation}
 	 which denotes that the disease will spread only if $\lambda>\frac{1}{m} \frac{2\mu \left\langle b\right\rangle }{\left\langle ab\right\rangle +\sqrt{\left\langle a^2\right\rangle \left\langle b^2\right\rangle }}$ \cite{Pozzana2017}. The epidemic threshold $\lambda_c$ depends on the properties of activity distribution $F(a)$ and attractiveness distribution $F(b)$. That is, the more heterogeneous the activity distribution and attractiveness distribution are, the easier it is for a disease to spread among population. The presence of positive correlations or uncorrelation between activity and attractiveness further facilitates contagion, while the presence of negative correlations hinders it \cite{Pozzana2017}. When activity distribution $F(a)$ and attractiveness distribution $F(b)$ are uncorrelated, Eq.~(\ref{eqn:ADAThreshold}) can be rewritten as
 	 	 \begin{equation}
 	 	~\label{eqn:ADAThresholdUncor}
 	 	\frac{\lambda}{\mu}=\frac{1}{m}\frac{2 \left\langle b\right\rangle }{\left\langle a\right\rangle\left\langle b\right\rangle +\sqrt{\left\langle a^2\right\rangle \left\langle b^2\right\rangle }}\text{.}
 	 \end{equation}
  
	 Here, we consider behavioral changes in the epidemic spreading,  including self-quarantine, self-protection, and social distancing. After getting infected, infected individuals will have less activity by factor $\omega$, i.e., $\omega a$, due to self-quarantine. Also, due to the self-protection of susceptible individuals, attractiveness of infected individuals will be reduced to $\eta b$. With social distancing, the number of generative links per activation for infected individuals is $c_m m$. Similarly, we can write the evolution of number of infected individuals in class $(a,b)$, which is given by:
	\begin{equation}
		\begin{split}
		\mathrm{d}_t I_{a,b} =&-\mu I_{a,b}^{t} + \lambda m \left( N_{a,b}^t-I_{a,b}^t\right) a \sum_{a^{'},b^{'}}\frac{\eta b^{'} I_{a^{'},b^{'}}^t}{N \left\langle b_{actual} \right\rangle}\\
		 &+ \lambda c_m m  \left( N_{a,b}^t-I_{a,b}^t\right) \frac{b}{N \left\langle b_{actual} \right\rangle} \sum_{a^{'},b^{'}} \omega a^{'} I_{a^{'},b^{'}}^t\text{,}
		 ~\label{eqn:IOrigin}
		 \end{split}
	\end{equation}
	where $\left\langle b_{actual} \right\rangle$ is the actually average attractiveness of all individuals in network. The first term in the right hand side describes infected individuals recover spontaneously with rate $\mu$. The second term represents susceptible individuals in class $(a,b)$ actively connect with infected individuals in class $(a^{'},b^{'})$ and get infected. The third term captures infected individuals in all possible classes $(a^{'},b^{'})$ actively connect with susceptible individuals in class $(a,b)$ and then infect them. 
	
	Considering the initial phases of epidemic spreading, when $I_{a,b}^t \ll N_{a,b}^t$, we have $I_{a,b}^t \approx 0$. Then, by summing up all possible classes $(a,b)$ on both sides of Eq.(\ref{eqn:IOrigin}), we get
	\begin{equation}
		\mathrm{d}_t I=-\mu I^t +\lambda m \frac{\left\langle a\right\rangle }{\left\langle b_{actual}\right\rangle } \eta \phi ^t + \lambda c_m  m \frac{\left\langle b\right\rangle }{\left\langle b_{actual}\right\rangle} \omega \psi^t \label{eqn:I}\text{,}
	\end{equation}
	where we set $\psi^t=\sum_{a,b}a I_{a,b}^t$ and $\phi^t=\sum_{a,b}b I_{a,b}^t$. In order to compute the epidemic threshold, auxiliary equations are needed. Multiplying both sides of Eq.(\ref{eqn:I}) by $a$ and $b$, respectively, and integrating across all classes:
	\begin{equation}
		\mathrm{d}_t \psi = -\mu \psi^t+ \lambda m \frac{\left\langle a^2 \right\rangle }{\left\langle b_{actual} \right\rangle} \eta \phi^t + \lambda c_m  m \frac{\left\langle ab \right\rangle}{\left\langle b_{actual} \right\rangle} \omega \psi^t\text{,}\\ \label{eqn:phi}
	\end{equation}

	\begin{equation}
		\mathrm{d}_t \phi = -\mu \phi^t + \lambda m \frac{\left\langle ab\right\rangle}{\left\langle b_{actual} \right\rangle} \eta \phi^t + \lambda c_m  m \frac{\left\langle b^2 \right\rangle}{\left\langle b_{actual} \right\rangle} \omega \psi^t\text{.} \label{eqn:psi}
	\end{equation} 

	 The Jacobian matrix $J$ of this set of linear differential equations (Eqs.(\ref{eqn:I})-(\ref{eqn:psi})) takes the form:\\
	 \begin{widetext}
	\begin{equation}      
	J=
	\left(                 
	\begin{array}{ccc}   
		
		-\mu & \lambda c_m  m \frac{\left\langle b \right\rangle }{\left\langle b_{actual} \right\rangle} \omega & \lambda m \frac{\left\langle a \right\rangle}{\left\langle b_{actual} \right\rangle}\eta\\ 
		
		0 & -\mu+\lambda c_m m \frac{\left\langle ab\right\rangle }{\left\langle b_{actual} \right\rangle}\omega & \lambda m \frac{\left\langle a^2 \right\rangle }{\left\langle b_{actual} \right\rangle}\eta\\
		
		0 & \lambda c_m m \frac{\left\langle b^2 \right\rangle}{\left\langle b_{actual} \right\rangle}\omega &
		-\mu+\lambda m  \frac{\left\langle ab\right\rangle }{\left\langle b_{actual}\right\rangle}\eta	
	\end{array}
	\right)\text{.}
	\label{eqn:NormalJacobian}                 
	\end{equation}
	
	Since the number of infected individuals can be negligible around the threshold, i.e. $I\approx0$. We can assume that $\left\langle b_{actual}\right\rangle\approx \left\langle b \right\rangle $. In addition, with the assumption that activity distribution $F(a)$ and attractiveness distribution $F(b)$ are uncorrelated, Eq.(\ref{eqn:NormalJacobian}) is simplified as:
		\begin{equation}       
		J=
		\left(                 
		\begin{array}{ccc}   
		-\mu & \lambda c_m m \omega & \lambda m \frac{\left\langle a \right\rangle }{\left\langle b \right\rangle }\eta\\	
		0 & -\mu+\lambda c_m m \left\langle a \right\rangle \omega & \lambda m \frac{\left\langle a^2\right\rangle }{\left\langle b\right\rangle}\eta\\
		0 & \lambda c_m m \frac{\left\langle b^2\right\rangle }{\left\langle b\right\rangle }\omega & -\mu+\lambda m \left\langle a\right\rangle \eta		
		\end{array}
		\right)\text{.}
		\label{eqn:SimpliedJacobian}                 
	\end{equation}	
	The disease will be able to spread only if the largest eigenvalue of the Jacobian matrix $J$ of the system Eq.(\ref{eqn:SimpliedJacobian}) is larger than zero. Thus, the condition for the presence of an endemic state is given by:

		\begin{align}
			~\label{eqn:Theshold}
			\lambda_c=\frac{1}{m}\frac{2\mu\left\langle b\right\rangle }{\sqrt{\left( c_m \omega \left\langle a \right\rangle\left\langle b \right\rangle - \eta\left\langle a \right\rangle\left\langle b \right\rangle\right) ^2+4 c_m \omega\eta\left\langle a^2 \right\rangle\left\langle b^2 \right\rangle} +\left( c_m \omega + \eta\right) \left\langle a \right\rangle \left\langle b \right\rangle}\text{.}
		\end{align}
		From Eq.(\ref{eqn:Theshold}), we can see that epidemic threshold $\lambda_c$ depends on the first ($\left\langle a\right\rangle $,$\left\langle b\right\rangle $) and second ($\left\langle a^2\right\rangle $, $\left\langle b^2\right\rangle $) moments of activity distribution $F(a)$ and attractiveness distribution $F(b)$, recovery rate $\mu$, activity factor $\omega$, attractiveness factor $\eta$ and connection factor $c_m$. Notably, when $\omega=\eta=c_m=1$, it recovers to the epidemic threshold (Eq. (\ref{eqn:ADAThresholdUncor})) of the original ADA models.
	\end{widetext}   			

	\section{Simulation Results on Synthetic Networks}~\label{sec:SimuonSyn}
	In this section, we perform numerical simulations to verify the theoretical results discussed in Section.\ref{sec:theory} and further investigate how the behavioral changes affect the epidemic dynamics. Since the observations on real networks indicate that the activity distribution $F(a)$ and attractiveness distribution $F(b)$ are heterogeneous \cite{Perra2012,Pozzana2017,Gozzi2020,Mancastroppa2020,Boccaletti2006,Newman2003}, we consider the heterogeneous condition where activity distribution follows $F(a)\sim a^{-\gamma_a}$ ($a\in[10^{-3},1]$, $\gamma_a=2.1$) and attractiveness distribution follows $F(b)\sim b^{-\gamma_b}$($b\in[10^{-3},1]$, $\gamma_b=2.1$). Unless otherwise specified, the default parameter values in our numerical simulations are set as $N=5000$, $\Delta T=1$, $I_0=50$, $\mu=0.1$ and $m=5$. $T=2000$ is chosen to guarantee the epidemic reaches asymptotic state. All experimental results are averaged 20 independent experiments.
	
	\subsection{Effects of Self-quarantine or Self-protection on the Epidemic}
	\begin{figure}
		\includegraphics[scale=0.2]{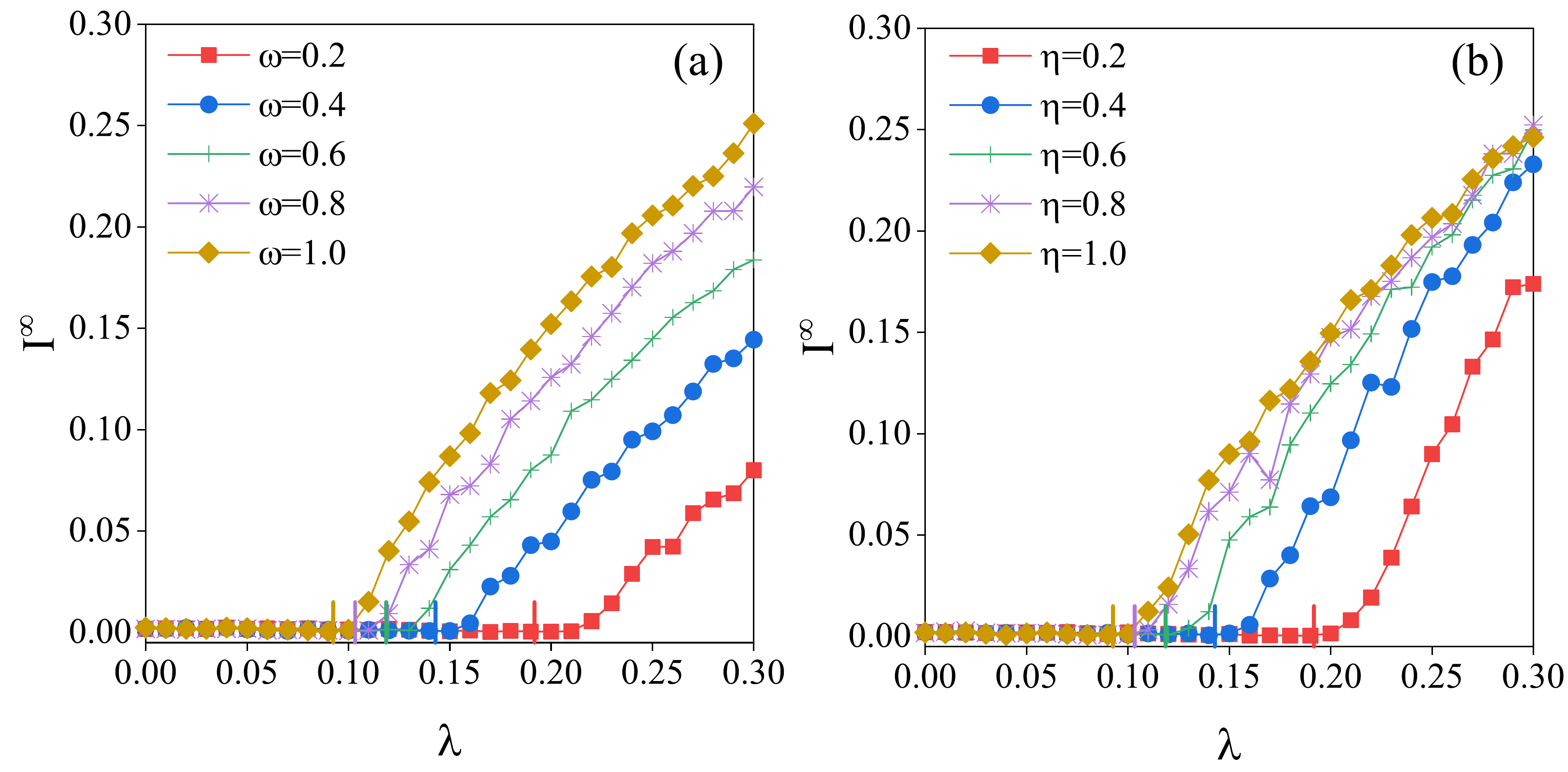}
		\caption{The epidemic size $I^\infty$ versus $\lambda$ under different behavioral changes. (a) Self-quarantine with $\omega$. $\eta=c_m=1$; (b) Self-protection with $\eta$. $\omega=c_m=1$. In both panels, vertical solid lines indicate the analytical thresholds derived from Eq.(\ref{eqn:Theshold})}
		\label{fig:omega_eta_single}
	\end{figure}
	
	To investigate the effects of self-quarantine factor $\omega$ or self-protection factor $\eta$ on the epidemic dynamics, we take representative $\omega$ or $\eta$ to perform the experiments, as shown in Fig.\ref{fig:omega_eta_single}. Obviously, there is a difference between the epidemic threshold derived from Eq.(\ref{eqn:Theshold}) and simulation results, since we assume that the number of infected individuals is ignored around the epidemic threshold. For the self-quarantine of infected individuals( Fig.\ref{fig:omega_eta_single}(a)), taking the SIS processes without behavioral changes(brown line) as baseline, we find that as the decrease of $\omega$, the epidemic threshold $\lambda_c$ is shifted towards higher values and the epidemic size $I^\infty$ decreases, which indicates that self-quarantine is effective both in reducing the epidemic size and delaying the epidemic. The role of self-protection of susceptible individuals controlled by parameter $\eta$ is also observed in Fig.\ref{fig:omega_eta_single}(b). Compared with self-quarantine, self-protection can not efficiently reduce the epidemic size. For instance, even for highly efficient self-protection factor with $\eta=0.2$, the epidemic size reaches to $0.17$. However, for self-quarantine factor with $\omega=0.2$, the epidemic size is reduced to $0.08$. This is because under the case of self-protection, susceptible individuals are more likely to contact with each other and form a local cluster. Once infected individuals actively connect with susceptible individuals, the disease can spread fast in the network. Therefore, experimental results show that both self-quarantine and self-protection can delay the epidemic by increasing the epidemic threshold, while only the former can decrease the epidemic size efficiently.
	\begin{figure}
		\includegraphics[scale=0.25]{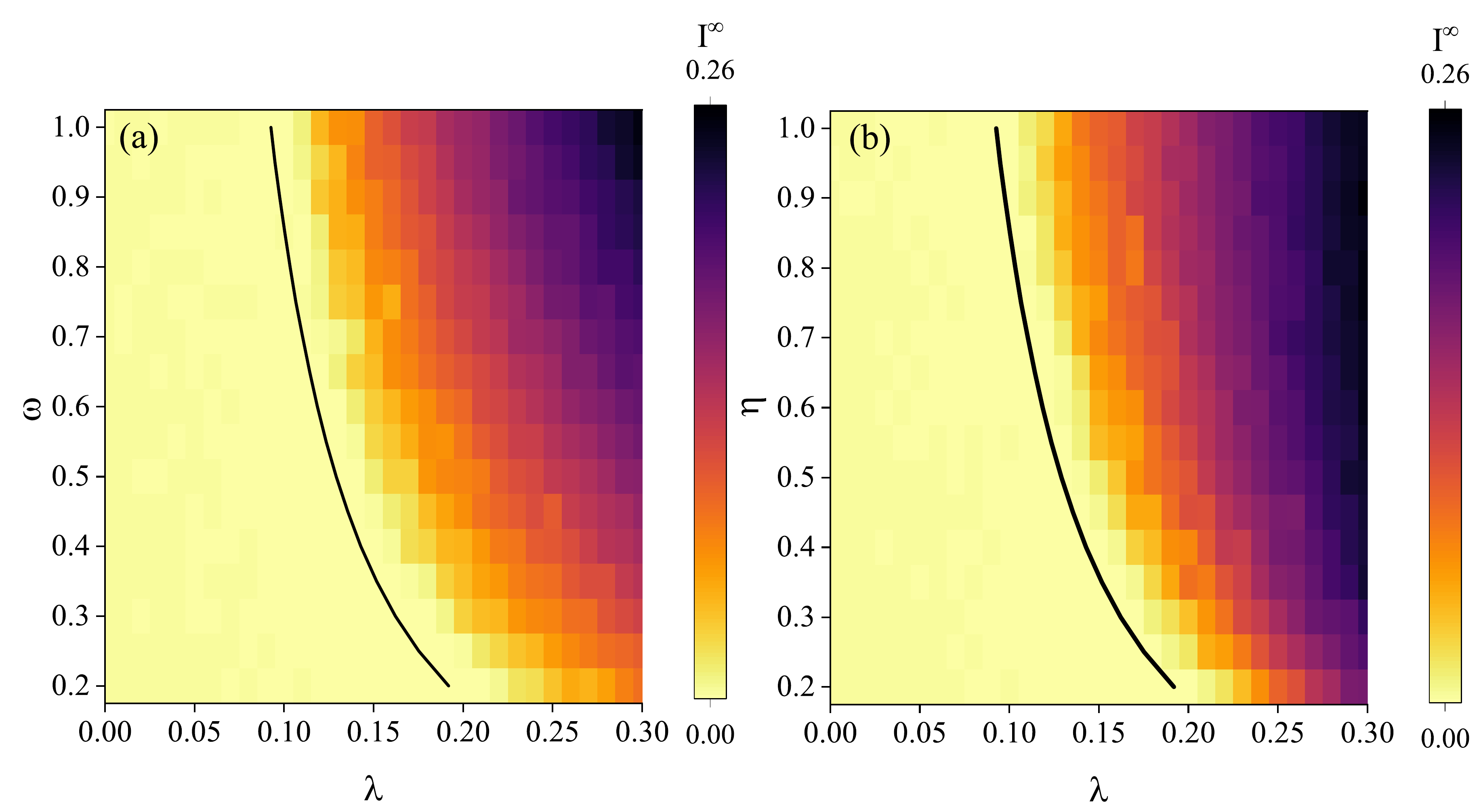}
		\caption{Effects of self-quarantine or self-protection on the epidemic. (a) Self-quarantine. The epidemic size $I^{\infty}$ versus $\lambda$ and $\omega$, $\eta=c_m=1$. (b) Self-protection. The epidemic size $I^{\infty}$ versus $\lambda$ and $\eta$, $\omega=c_m=1$. In both panels, the black solid lines indicate the analytical threshold derived from Eq.(\ref{eqn:Theshold}).}
		\label{fig:omegaOReta_heat}
	\end{figure}

	In order to further explore the effects of self-quarantine of infected individuals or self-protection of susceptible individuals on the epidemic threshold, we observe the epidemic size $I^\infty$ at the steady state versus different infection rate $\lambda$ and self-quarantine factor $\omega$ or self-protection factor $\eta$, as shown in Fig.\ref{fig:omegaOReta_heat}. We find that the epidemic thresholds derived from Eq.(\ref{eqn:Theshold}) are consistent with Monte Carlo simulation results. In Fig.\ref{fig:omegaOReta_heat}(a), as the decrease of $\omega$, the epidemic threshold $\lambda_c$ increases, which indicates that the implementation of self-quarantine could delay the epidemic. The same role of self-protection is also observed in Fig.\ref{fig:omegaOReta_heat}(b). In addition, substituting $c_m=1$ into Eq.(\ref{eqn:Theshold}), we can get the same expression by swapping self-quarantine factor $\omega$ and self-protection factor $\eta$. Thus, self-quarantine factor $\omega$ and self-protection factor $\eta$ with same parameter values have the same epidemic threshold, which demonstrates that self-quarantine of infected individuals and self-protection of susceptible individuals could delay the epidemic to the same time when implemented with the same strength.

	\begin{figure}
		\includegraphics[scale=0.12]{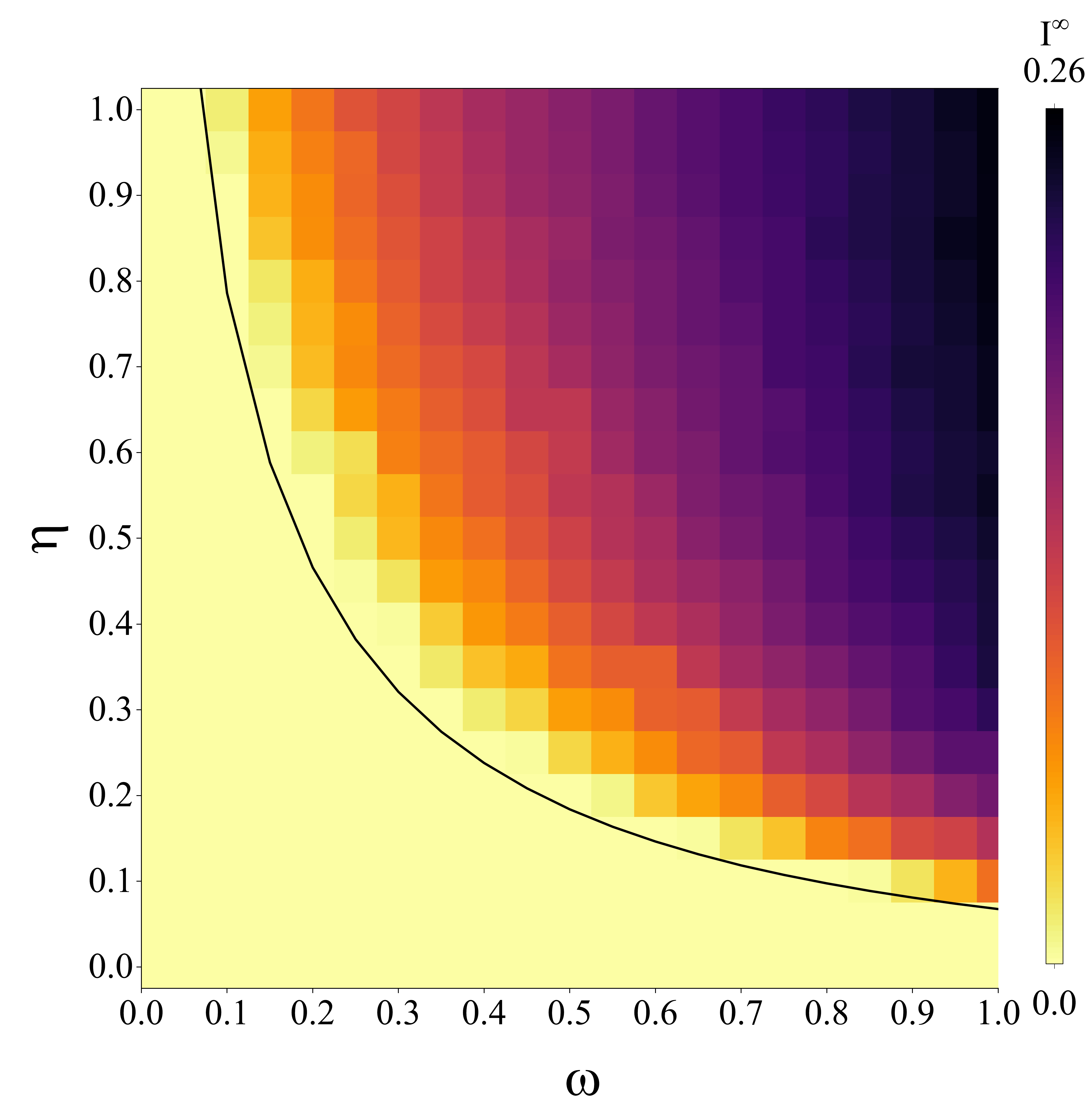}
		\caption{Effects of joint changes of self-quarantine $\omega$ and self-protection $\eta$ on the epidemic dynamics, $\lambda=0.3$. The black solid line indicates the analytical threshold derived from Eq.(\ref{eqn:Theshold}).}
		\label{fig:omegaANDeta_heat}
	\end{figure}

	Finally, we explore the impacts of joint changes of self-quarantine and self-protection on the epidemic threshold and epidemic size, as shown in Fig.\ref{fig:omegaANDeta_heat}. We observe  that the epidemic threshold derived by our theoretical model approximates well with Monte Carlo simulation results, providing a strong evidence to our model. The black solid line splits the parameter space in two parts: to its left, the joint effects of $\omega$ and $\eta$ result in the disappearance of epidemic from the populations. While to its right, the joint effects of self-quarantine and self-protection are not sufficient enough to hinder the epidemic spreading. In addition, introducing increasingly stronger self-quarantine and self-protection(i.e. as $\omega$ and $\eta$ decrease), $I^\infty$ becomes progressively smaller. The reason is that self-quarantine of infected individuals reduces the chance of contacts between them and susceptible individuals, and self-protection of susceptible individuals ensures they are more likely to contact with each other. Experimental results show that joint changes of self-quarantine and self-protection can increase the epidemic threshold $\lambda_c$ and decrease epidemic size $I^\infty$.

	\subsection{Effect of Social Distancing on the Epidemic}
	\begin{figure}
		\includegraphics[scale=0.4]{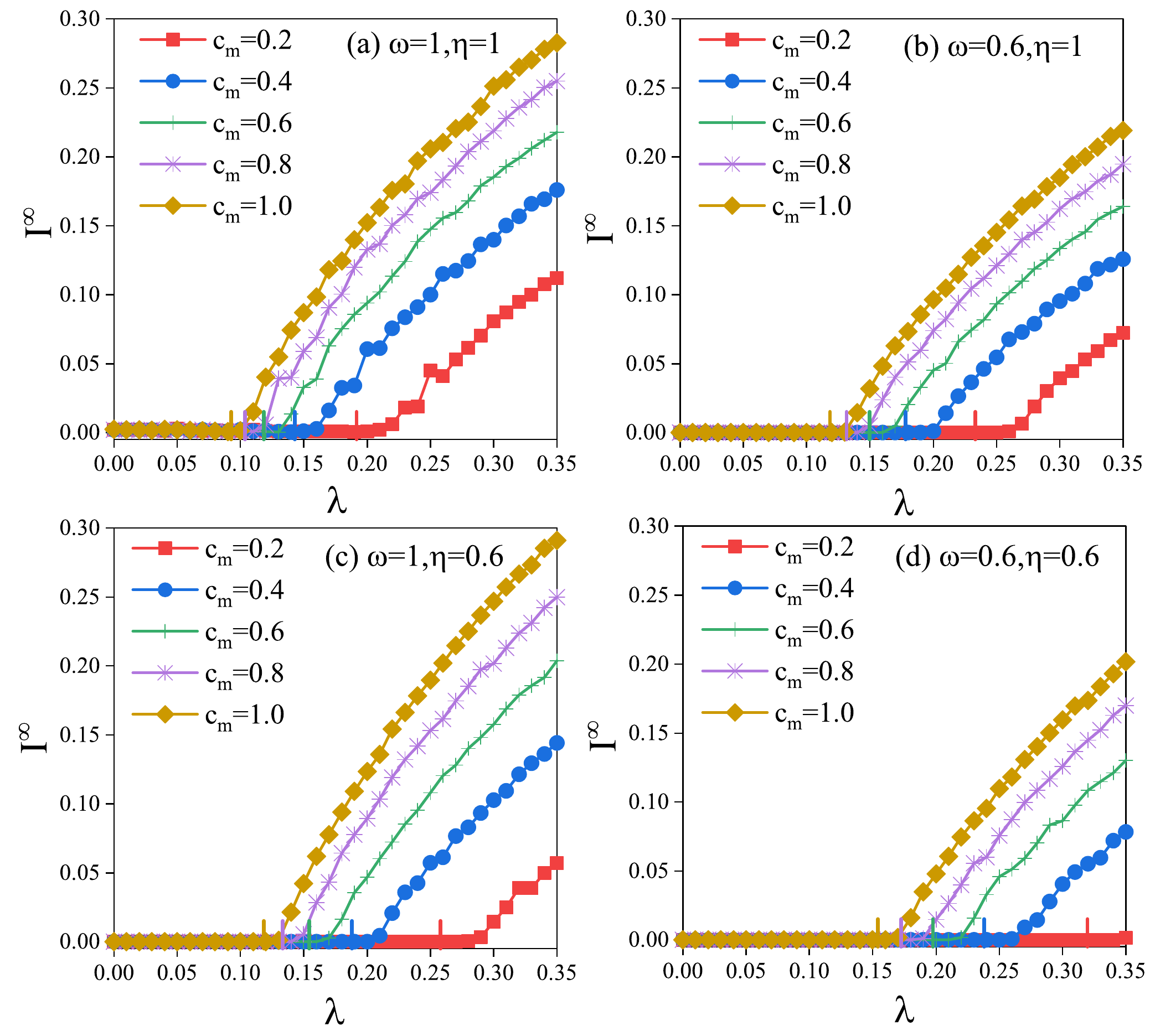}
		\caption{The final epidemic size $I^\infty$ as a function of $\lambda$ under the different joint changes of behavior. (a) Social distancing under different factor $c_m$. $\omega=1$, $\eta=1$; (b) Self-quarantine($\omega$) and social distancing($c_m$). $\omega=0.6$, $\eta=1$; (c) Self-protection($\eta$) and social distancing($c_m$). $\omega=1$, $\eta=0.6$; (d) Self-quarantine($\omega$), self-protection($\eta$) and social distancing($c_m$). $\omega=0.6$, $\eta=0.6$. In all panels, vertical solid lines indicate the analytical threshold derived from Eq.(\ref{eqn:Theshold}).}
		\label{fig:socialdistancing}
	\end{figure}
	To explore the effects of social distancing of infected individuals on the epidemic dynamics, we carry out simulations under different factors $c_m$, as shown in Fig.\ref{fig:socialdistancing}(a). As the enforcement of social distancing(decrease of $c_m$), the epidemic threshold is increased to higher values and the epidemic size decreases gradually, which indicates that social distancing could delay the epidemic and reduce the epidemic size. Obviously, the reason is that social distancing of infected individuals reduces the number of contact with others, and thus their effective infection rate is reduced.
	
	Next, with the introduction of self-quarantine and self-protection, we analyze the joint changes of self-quarantine and social distancing, self-protection and social distancing, respectively, as shown in Fig.\ref{fig:socialdistancing}(b)(c). Compared with the solely social distancing(Fig.\ref{fig:socialdistancing}(a)), the introduction of self-quarantine further delays the epidemic and reduces the epidemic size as well (Fig.\ref{fig:socialdistancing}(b)). The joint changes of self-protection and social distancing also delay the epidemic by increasing the epidemic threshold, but could not reduce the epidemic size effectively, as shown in Fig.\ref{fig:socialdistancing}(c). Interestingly, compared with the simultaneous implementation of self-quarantine and social distancing , the joint changes of self-protection and social distancing have the larger epidemic threshold, which implies that the latter can more effectively delay the epidemic. As an example, for self-quarantine and social distancing with $c_m=0.2$ (Fig.\ref{fig:socialdistancing}(b), red line), the epidemic threshold reaches to 0.23. However, for self-protection and social distancing with $c_m=0.2$ (Fig.\ref{fig:socialdistancing}(c), red line), the epidemic threshold is 0.26. The reason for this phenomenon is that the self-quarantine of infected individuals reduces the chance to attend social activities, so that the implements of social distancing have less impact on epidemic. 
	
	Finally, we investigate the joint changes of self-quarantine, self-protection and social distancing in Fig.\ref{fig:socialdistancing}(d). We observe that compared with the other three kinds of behavioral combinations (Fig.\ref{fig:socialdistancing}(a)-(c)), the joint changes of self-quarantine, self-protection and social distancing will result in the largest epidemic threshold and the lowest epidemic size. This is because self-quarantine and social distancing of infected individuals reduce the chance and number of contacts between them and other individuals, and self-protection of susceptible individuals guarantee they are more likely to contact with each other. Based on the above analysis, individual behaviors should be implemented simultaneously to curb the epidemic.
	
	\section{Simulation Results on Real Networks}~\label{sec:SimuonReal}
	To further investigate how change in individual behavior affect epidemic dynamics on real networks, we use the data set during the Infectious SocioPatterns event(INFECTIOUS: STAY AWAY). The data set contain 10999 contact information between 318 participants and the corresponding time-stamped information, where each node represents a participant and each time-resolved link denotes the contact between participants. The data set is available on the website \url{http://www.sociopatterns.org/datasets/}.
	
\begin{figure}
	 	\includegraphics[scale=0.35]{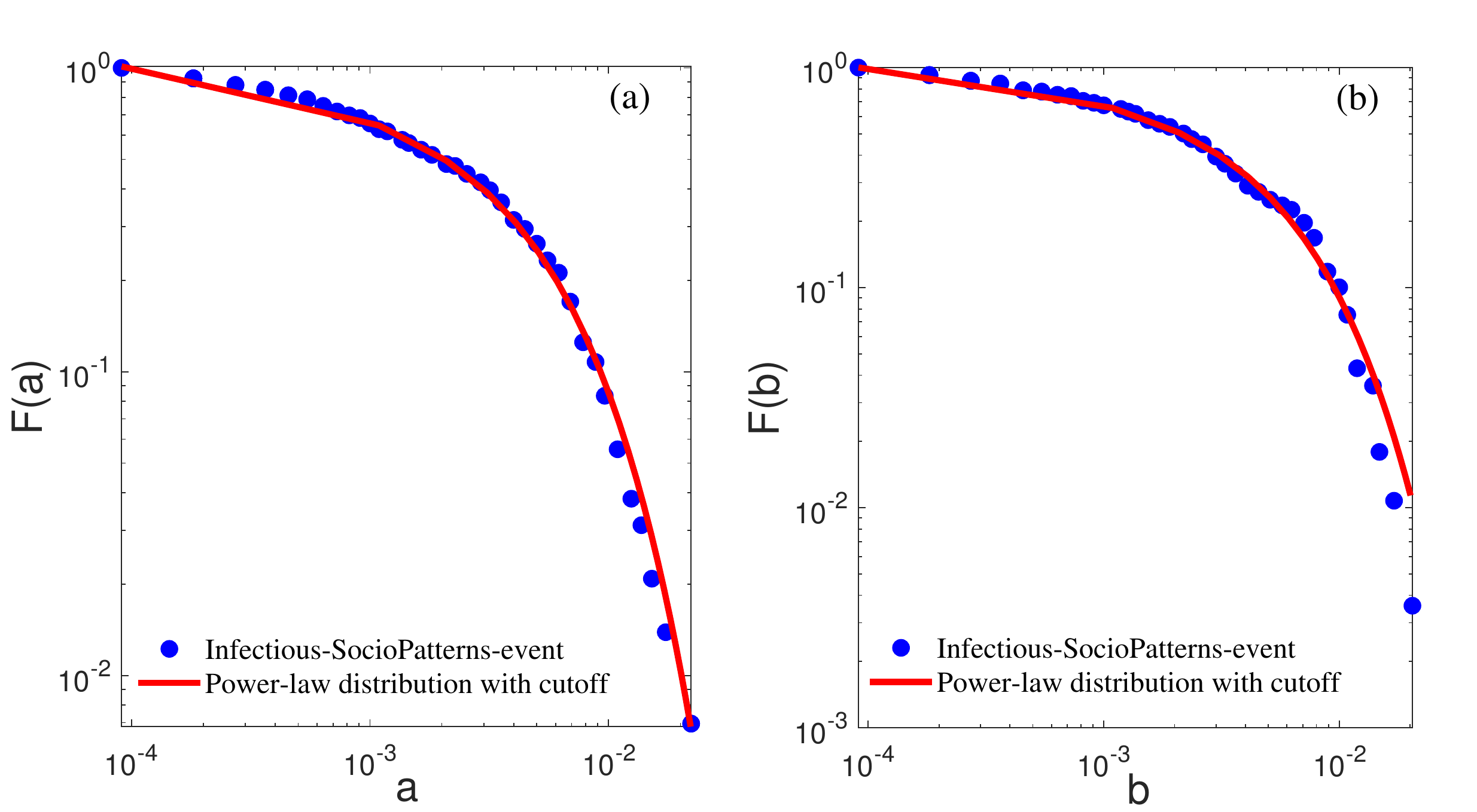}
	 	\caption{The activity distribution $F(a)$ and attractiveness distribution $F(b)$ in the Infectious-SocioPattern-event data set. The blue filled circles represent the real data and the red curves represent the fitting result of power-law distribution with cutoff, $y(x)=\alpha x^{-\gamma}e^{-\beta x}$ with $95\%$ confidence bounds. (a) The fitting result of activity distribution $F(a)$, $\alpha=0.4126$, $\gamma=0.09832$, $\beta=202$.  (b) The fitting result of attractiveness distribution $F(b)$, $\alpha=0.4403$, $\gamma=0.09024$, $\beta=199$.}
	 	\label{fig:actandattforRealData}
	 \end{figure}

	 In order to characterize an individual's activity and attractiveness in this network, we measure the activity $a_i$ defined as the fraction of active contacts of node $i$ per unit time, which is computed as $a_i=\frac{e_{i,out}}{\sum_j e_{j,out}}$, where $e_{i,out}$ is the out-strength of node $i$ integrated across the entire time slices \cite{Alessandretti2017}. Similarly, attractiveness $b_i$ is defined as the fraction of passive contacts of node $i$ per unit time, which is computed as $b_i=\frac{e_{i,in}}{\sum_j e_{j,in}}$ , where $e_{i,in}$ is the in-strength of node $i$ integrated across the entire time slices. We depict the distribution of activity and attractiveness in the Infectious-SocioPatterns-event data set, as shown in Fig.~\ref{fig:actandattforRealData}. Since usually the activity distribution $F(a)$ and attractiveness distribution $F(b)$ are often truncated in real networks, we use the power law distribution with cutoff to fit them \cite{Clauset2009}. The fitting results show that the activity distribution $F(a)$ and attractiveness distribution $F(b)$ are both less heterogeneous.  
	 
	 Since the finite duration of real data is not sufficient to allow the epidemic dynamics to reach asymptotic state, we consider the sequence replication method\cite{Starnini2012} to solve this difficulty. It makes the contact data repeated periodically, defining a extended function such that $\chi^{Rep}(i,j,t)=\chi(i,j,t$ mod $T)$. We set the duration of new contact data to 5 times the duration of original contact data $T$.
	 \begin{figure}
	 	\includegraphics[scale=0.25]{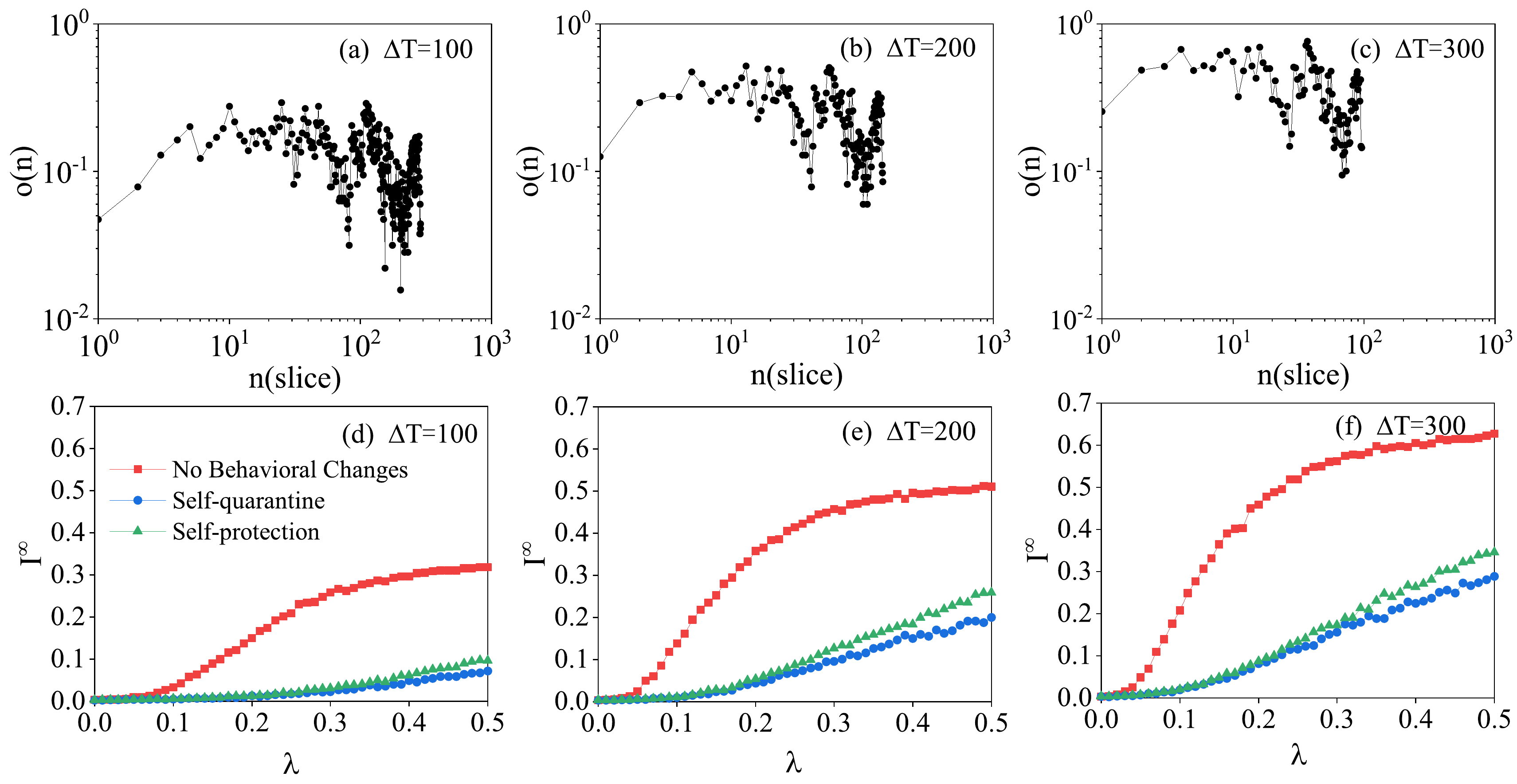}
	 	\caption{The fraction of active nodes O(n) and epidemic size $I^{\infty}$ in Infectious-SocioPattern-event data set for different time slice $\Delta T$. Different color lines indicate different behavioral changes: No behavior changes(green line), Self-quarantine(red line) and Self-protection(blue line). (a)(d) $\Delta T=100$; (b)(e) $\Delta T=200$; (c)(f) $\Delta T=300$. The parameter is set as $\mu=0.01$. Each point is averaged over 200 independent simulations with $1\%$ random infection seeds.} 
	 	\label{fig:timesliceandepidemic}
	 \end{figure}
 
 	To explore how choice of time slices $\Delta T$ affects the epidemic dynamics, we divide the original data according to the time slices $\Delta T=100,200,300$, respectively, as shown in Fig.\ref{fig:timesliceandepidemic}. As the increase of time slices, the fraction of active nodes in a time slices increases, which reflects the earlier time that nodes participate in interaction. Since nodes appearing at earlier time will lead to stronger infectious ability, which is called aging effects\cite{Cardillo2014}, the epidemic is easier to spread.
 	
 	Next, we compare the different individual behavior changes versus infection probability $\lambda$ in Fig.~\ref{fig:timesliceandepidemic}(d)-(f). For self-quarantine of infected individuals, in order to reflect the reduction in activity of infected node $i$, we eliminate the edges whose ``source`` node $i$ is infected. For self-protection of susceptible individuals, in order to reflect the reduction in attractiveness of infected node $i$, the edges whose ``target`` node $i$ is infected are eliminated. Because we have assumed that the implementation of self-quarantine of infected individuals eliminate the edges whose ``source'' node is infected in every time slices $\Delta T$,  we don't need to consider social distancing in real network simulation. As expected, compared with no behavioral changes, self-quarantine and self-protection delay the epidemic spreading by increasing the epidemic threshold. In addition, self-quarantine can further lower the epidemic size, which are consistent with our results in synthetic networks.
 	
	 \begin{figure}
	 	\includegraphics[scale=0.8]{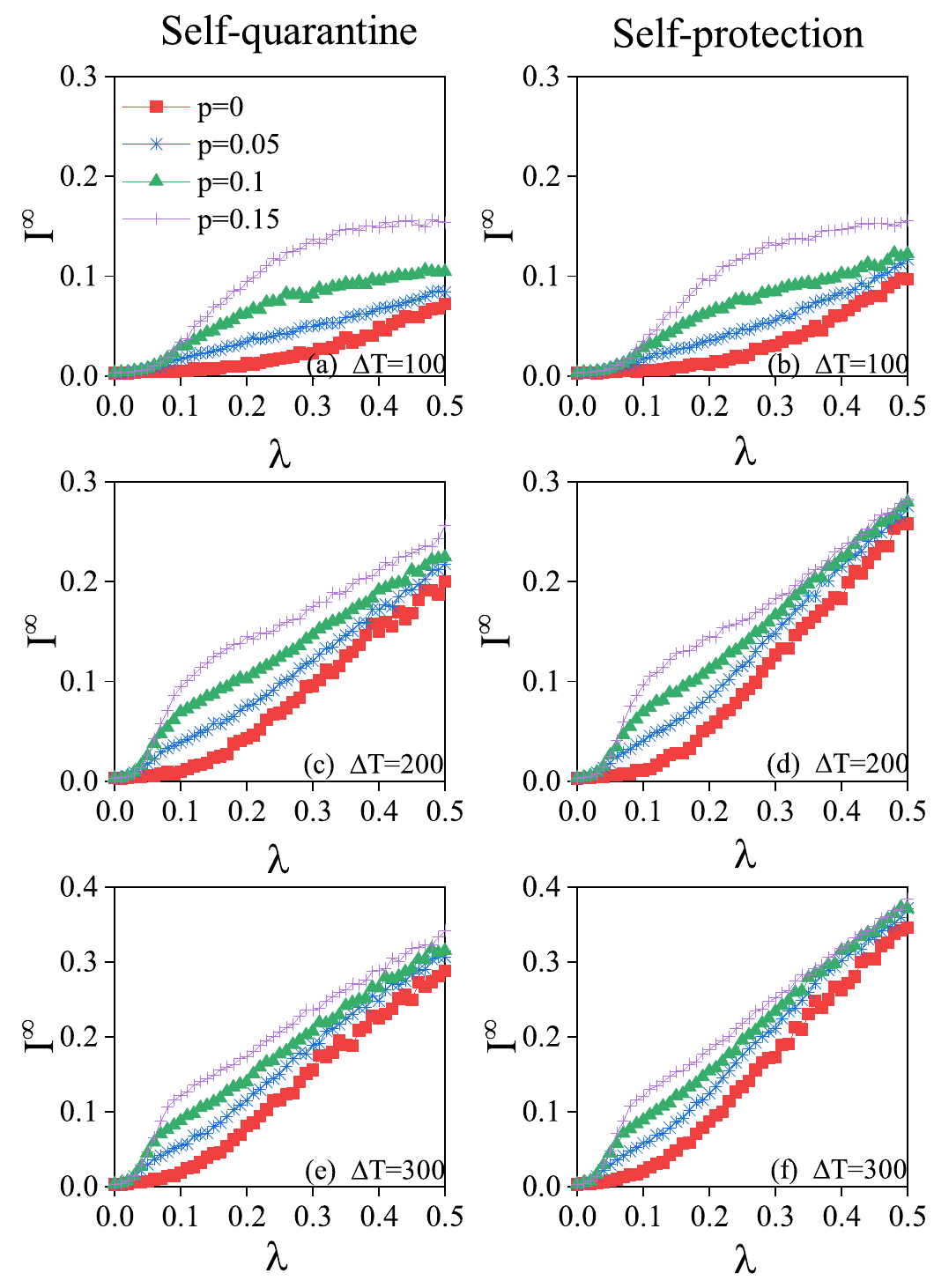}
	 	\caption{Effects of timing in behavioral changes of self-quarantine(left column) and self-protection(right column). (a),(b) $\Delta T=100$; (c),(d)$\Delta T=200$; (e),(f)$\Delta T=300$. Each panel shows the epidemic size $I^{\infty}$ as a function of $\lambda$. The parameter is set as $\mu=0.01$. Each point is averaged over 200 independent simulations with $1\%$ random infection seeds.} 
	 	\label{fig:epidemicunderdifferenTime}
	 \end{figure}
	Further, since the different ability of individuals to perceive risk, behavioral changes are implemented at different time. We assume that when the fraction of infected individuals among population reaches to $p$, individuals are aware of infected risk and change their behavior. We consider the time slices $\Delta T$=100, 200 and 300, respectively. As shown in Fig.\ref{fig:epidemicunderdifferenTime}, simulation results show that the earlier behavioral changes are implemented, the lower epidemic size will be. For instance, in Fig.\ref{fig:epidemicunderdifferenTime}(a), we observe that when self-quarantine is implemented immediately at $p=0$, the epidemic size is about 0.08. As the delay of timing in behavioral changes implements, e.g. $p=0.15$, the epidemic size reaches to 0.15. However, as the increase of time slices $\Delta T$(Fig.\ref{fig:epidemicunderdifferenTime}(a)(c)(e)), the difference between epidemic size decreases gradually under different implementation time. As an example, for time slices $\Delta T=100$, the epidemic size is 0.08 and 0.15, respectively, when self-quarantine is implemented at $p=0$ and $p=0.15$. As the increase of $\Delta T$, e.g. $\Delta T=300$, the corresponding epidemic size is 0.29 and 0.32. The reason for this phenomenon is that due to the aging effects, the increase of time slices $\Delta T$ will lead to stronger propagation ability of nodes. Although behavioral changes are implemented at different time, the propagation ability of nodes is so strong that it weakens the role of timing that individuals change their behavior. Similar results can be observed in the analysis of the interaction between timing and self-protection, see Fig. \ref{fig:epidemicunderdifferenTime}(b)(d)(f).	
	
	\section{Conclusion And Discussion}~\label{sec.conclu}
	 In this work, we model three typical behavioral changes driven by infected individuals, which play a fundamental role in altering the epidemic, that is, self-quarantine and social distancing of infected individuals, self-protection of susceptible individuals. By connecting the relationship between individual behavioral changes and their social attributes, we investigate the issue on the ADA model. Specifically, self-quarantine and social distancing of infected individuals reduce their activity and the number of generative links per activation, respectively. Self-protection of susceptible individuals leads to the reduction in attractiveness of infected individuals. In this way, we could investigate the joint changes of behavior. Then we adopt the mean field theory and Monte Carlo experiments to analyze how behavioral changes affect the threshold and prevalence of epidemic.
	
	It is found that self-quarantine and social distancing of infected individuals, self-protection of susceptible individuals can delay the epidemic by increasing the epidemic threshold, but only self-quarantine and social distancing of infected individuals can decrease the epidemic size. Next, joint changes of behavior will further delay epidemic and decrease epidemic size. Lastly, the simulation results in real networks also reveal that self-quarantine of infected individuals and self-protection of susceptible individuals both increase epidemic threshold, while the former could decrease epidemic size efficiently. Besides, the earlier behavioral changes are implemented, the lower epidemic size will be. These results will be helpful for understanding the interaction of individual behaviors in the epidemic process and also help design reasonable intervention strategies for the control of the epidemic.
	
	Of course, there are some limitations to this work. Firstly, in order to theoretically derive the epidemic threshold, we ignore the correlations between activity and attractiveness, which should not be simplified as a precise representation of real world. Secondly, although we focused on some fundamental features of realistic world(e.g. heterogeneity of contacts, activity and attractiveness for individuals, behavioral changes), we neglect some other important features, such as an age-structured populations, the preferential contacts among individuals. The above features represent the possibly future challenges for this area.
	\begin{acknowledgments}
	This work was supported by  the National Natural Science Foundation of China under Grant No. 91746203, the National Key Research and Development Program of China under Grants No. 2017YFE0117500, Natural Science Foundation of Shanghai under Grant No. 20ZR1419000. 
	\end{acknowledgments}
	\bibliographystyle{unsrt} 
	\bibliography{ref}
\end{document}